\documentclass[useAMS,usenatbib]{mn2e}

\usepackage{graphicx}
\usepackage{txfonts}
\usepackage{natbib}
\title[]{Spatially resolved LMC star formation history: I. Outside in evolution of the outer LMC disk.}
\author[I.\, Meschin et al.]
{I..\, Meschin,
C. Gallart,$^{1,2,3}$\thanks{email: carme@iac.es} A. Aparicio,$^{1,2}$ S.L. Hidalgo,$^{1,2}$   M. Monelli,$^{1,2}$
\newauthor
P.B. Stetson, $^{4}$ R. Carrera,$^{1,2}$\\
$^{1}$ Instituto de Astrof\'{i}sica de Canarias, Calle V\'\i a L\'actea s/n, 38205 La Laguna, Tenerife, Spain \\
$^{2}$ Departamento de Astrof\'{i}sica, Universidad de La Laguna,  E-38206 La Laguna, Tenerife, Spain\\
$^{3}$ Visiting Astronomer, Cerro Tololo Inter-American Observatory.
CTIO is operated by AURA, Inc.\ under contract to the National Science
Foundation.\\
$^{4}$ NRC Herzberg Institute for Astrophysics, 5071 West Saanich Road, Victoria, BC V9E 2E7, Canada\\}

\begin{document}

\date{Draft Version November 11, 2013}


\maketitle

\label{firstpage}

\begin{abstract}  
We study the evolution of three fields in the outer LMC disk (R$_{gc}$=3.5--6.2 Kpc). Their star formation history indicates a stellar populations gradient such that younger stellar populations are more centrally concentrated. We identify two main star forming epochs, separated by a period of lower activity between $\simeq$7 and $\simeq$4 Gyr ago. Their relative importance varies from a similar amount of stars formed in the two epochs in the innermost field, to only 40\% of the stars formed in the more recent epoch in the outermost field.  The {\it young} star forming epoch continues to the present time in the innermost field, but lasted only till $\simeq$0.8 and 1.3 Gyr ago at R$_{gc}$=5.5\degr\ and 7.1\degr\, respectively. This gradient is correlated with the measured HI column density and implies an outside-in quenching of the star formation, possibly related to a variation of the size of the HI disk. This could either result from gas depletion due to star formation or ram-pressure stripping, or from to the compression of the gas disk as ram-pressure from the Milky Way halo acted on the LMC interstellar medium. The latter two situations may have occurred when the LMC first approached the Milky Way.
\end{abstract}


\section{Introduction}\label{sec:intro}

The Large Magellanic Cloud (LMC) possesses some characteristics that makes it an ideal laboratory for the study of galaxy formation and evolution. (1) Its closeness makes the oldest main sequence turnoff (oMSTO) at V =22.5, which observation is essential for a reliable determination of the full SFH, accessible from the ground. This is important to allow wide field observations to be carried on, through which, with survey imaging capabilities, it should be possible to image a substantial  fraction (if not all) of the galaxy. (2)  It contains star clusters in wide age ranges (from old globular clusters to --after the so-called age-gap-- a large population of intermediate-age star clusters  and regions of active star formation) and field stars formed over its whole lifetime. This will allow us to compare the SFH derived from star clusters and field stars, providing a local test bed of the use of star clusters as a proxy of the SFH of distant galaxies. (3) It hosts the only resolved bar in the Local Group, and thus offers the possibility of studying the mixing effects of a bar in a stellar population, in addition to the formation and evolution of the bar itself; and (4) with the SMC and the Milky Way, it is part of a currently interacting system, and therefore, it can provide important insight on the effects of interactions in galaxy evolution. 

In spite of this, a complete picture of the LMC SFH, extending to the first events of star formation, and taking into account its prominent spatial gradients \citep{Gallart08LMC} is still sorely lacking. The vastness of the photometrically accessible LMC stellar populations, and its huge angular extent, makes of such project a very ambitious one, which can be approached only with wide field imaging and spectroscopic facilities. In this paper, we present  a determination of the SFH in three 36\arcmin$\times$36\arcmin\ fields in the outer LMC disk (R$_{gc}\geq$ 4.0\degr\, or 3.5 Kpc) for which we have obtained CMDs reaching the oMSTOs with good photometric accuracy and high completeness. These CMDs are populated and deep enough to ensure a robust determination of their SFHs, which will allow us to explore the presence of  SFH variations across the LMC disk and and to characterise them. The same data have been previously used to explore the extent of the LMC disk and to discuss the stellar populations present at a radial distance of $\simeq$ 6 Kpc \citep{Gallart04LMC} and to argue for an outside-in disk evolution in the LMC \citep{Gallart08LMC}. It has also been the subject of spectroscopic follow-up by \citet{Carrera08LMC}, who used observations of the CaII triplet on RGB stars to demonstrate that, even though a metallicity gradient exist in the outer LMC disc, in the sense that the mean metallicity decreases outwards, the age-metallicity relation remains the same in all three fields.

This paper is the first of a series aimed at piecing together a reliable picture of the spatial and temporal variations of the LMC SFH.  It is organised as follows: section~\ref{observations} presents the observations, photometry and artificial stars tests.  Section~\ref{sec_cmds} introduces the CMDs of the three LMC fields and discusses their main features. Section~\ref{derivation} discusses the procedure of SFH derivation, and details the parameters adopted in this particular case. Section~\ref{sfhs} presents the resulting SFHs for the three LMC fields. Finally, Sections~\ref{prev} and~\ref{discussion} discuss these SFHs in the context of former results regarding field and cluster data, and their implications in relation to the study of galaxy formation and evolution. 

\section{Observations, photometry and crowding tests} \label{observations}

We obtained V and I images of three fields in the LMC, on November 30th and December 1st and 3rd, 1999, and on January 17th and 18th, 2001, using the Mosaic II CCD Imager on the CTIO Blanco 4m telescope. The Mosaic II was composed by 8 SITe 2K$\times$4K CCDs, which provided a total field of 36\arcmin $\times$ 36\arcmin, with a resolution of 0.27\arcsec/pixel. These fields were chosen to span a wide range of galactocentric distances, from $\sim$ 4.0\degr\ to 7.1\degr\ from the centre of the LMC to the North.  The coordinates of the three fields and the data obtained for each of them are detailed on Table~\ref{fields}. In \citet{Gallart08LMC}, we named the fields according to their right ascension and declination (J2000.0) as LMC0514-6503, LMC0513-6333 and LMC0513-6159. In this paper, for easier reference we will use the short names LMC2, LMC1 and LMC0, for the three fields in order of increasing galactocentric distance. Seeing was typically around 1.0\arcsec, except during the observations of field LMC1, when it went up to 1.4\arcsec\ on average. The first two nights of the 1999 run and both nights of the 2001 run were photometric, and the \citet{Landolt92} standard star fields SA92, SA95, SA98, SA104 were observed several times each night for calibration purposes.

\begin{table*}
\centering
\begin{minipage}{160mm}
\caption{Data obtained in the LMC}
\begin{tabular}{rrrrrrrrrr}
\hline
Field & Short name &$\alpha_{2000}$& $\delta_{2000}$ & R(\degr)\footnote{Distance from the LMC dynamical center: $\alpha_{2000}=$ 05:17:36, $\delta_{2000}=$ -69:02 \citep{Kim98}. Note than different centres were used as a reference in \citet{Gallart04LMC}} & R(Kpc) & Exposures V (sec) & Exposures I (sec) & seeing & E(B-V)\footnote{For the two outermost fields, the reddening values given by \citet{Schlegel98} have been used. For the innermost field, mean reddening values have been estimated by requiring a good fit of the CMD by the same isochrones as for the outermost fields (see Figure~\ref{cmds}. A$_V$/E(B-V)=3.315 and A$_I$/E(B-V)=1.940 have been adopted following the same reference} \\
\hline

LMC0514-6503 & LMC2 & 05:13:43 & -65:03:20 & 4.0 & 3.5 &2+6+60+500+2$\times$600 &2+6+60+4$\times$600+800 &  1\arcsec & 0.050\\
LMC0513-6333& LMC1 & 05:13:13 & -63:33:20 & 5.5 & 4.8 &6+60+5$\times$800 & 6+60+10$\times$600 & 1.4\arcsec & 0.037\\
LMC0513-6159 & LMC0 & 05:13:17 & -61:58:40 & 7.1 & 6.2 &2+3$\times$6+3$\times$15+60+6$\times$900& 2+2$\times$6+10+60+8$\times$600& 1\arcsec & 0.026\\
\hline

\label{fields}
\end{tabular}
\end{minipage}
\end{table*}

The Mosaic frames were reduced in a standard way, as indicated in the {\it CTIO Mosaic II Imager User Manual} \citep{Armandroff00}, using the MSCRED package within IRAF. 
Profile-fitting photometry of the LMC fields was obtained using the DAOPHOT/ALLFRAME suite of computer programs \citep{Stetson87, Stetson94}. Typically 150-200 PSF stars were selected in each chip in order to accurately model the PSF and its spatial variations. A Moffat function with $\beta$=2.5, and a PSF varying quadratically across the image were typically used. The ALLFRAME input star list was obtained by finding stars in all individual images, matching them and keeping all those that were found in at least two images. The final photometry list was computed by DAOMASTER from the combination of  ALLFRAME measurements for each star.  Stars with $\sigma_V$ and $\sigma_I$ below 0.1,  CHI$<$8 ($<$4 for field LMC0) and {SHARP}$<$1 ($<$0.5 for LMC0), as computed by DAOMASTER were kept for the final photometric catalogue. This resulted in a total of $\simeq$ 72000, 114000 and 290000 stars in fields LMC0, LMC1 and LMC2 respectively.  

 The profile fitting photometry was then referred to a system of synthetic-aperture photometry by the method of growth-curve analysis \citep{Stetson90}. Calibration of these instrumental data to the photometric system of \citet{Landolt92} was carried out as described by \citet{Stetson00,Stetson05}.

We carried on a large number of artificial stars tests to characterise the completeness and uncertainties of our photometry, with a similar strategy as that described in \citet{Monelli10sfhcetus} and \citet[]{Gallart96a}. Almost 4 million artificial stars per field were injected in the images, in the following way. We performed 30 iterations for LMC0 and 32 for LMC1 and LMC2 with 15520 artificial stars per chip, distributed in a regular grid of equilateral triangles, and separated by a distance R=(rad$_{PSF}$+rad$_{fit}$ + 1). This avoids the wings of the artificial stars to overlap with the core of the PSF used for the fit. Gallart et al. (1999) empirically determined that no overcrowding effects were measurable even when somewhat smaller distances were used. In each iteration, the grid was shifted by a few pixels. The artificial stars were distributed in the colour-magnitude plane so that the full range of luminosity and colour of the observed CMD was covered, by using two synthetic CMDs calculated using IAC-star \citep{iacstar}. One of them covered the full range of magnitudes observed, down to M$_I$=+5. In this one, over 60\% of the stars were fainter than M$_I$=3.5. In order to better sample the bright part of the CMD, a second synthetic CMD, with magnitude limit M$_I$=+3.5 was used.  The recovered and injected magnitudes of the artificial stars, together with the information on the stars that were lost, were recorded in a {\it crowding trial table} \citep[see][]{Aparicio95}, that was afterwards used to simulate the observational effects on the synthetic CMDs (see Section~\ref{derivation}). 

\section{The Colour Magnitude Diagrams} \label{sec_cmds}

\begin{figure*}
\centering
\includegraphics[width=16cm]{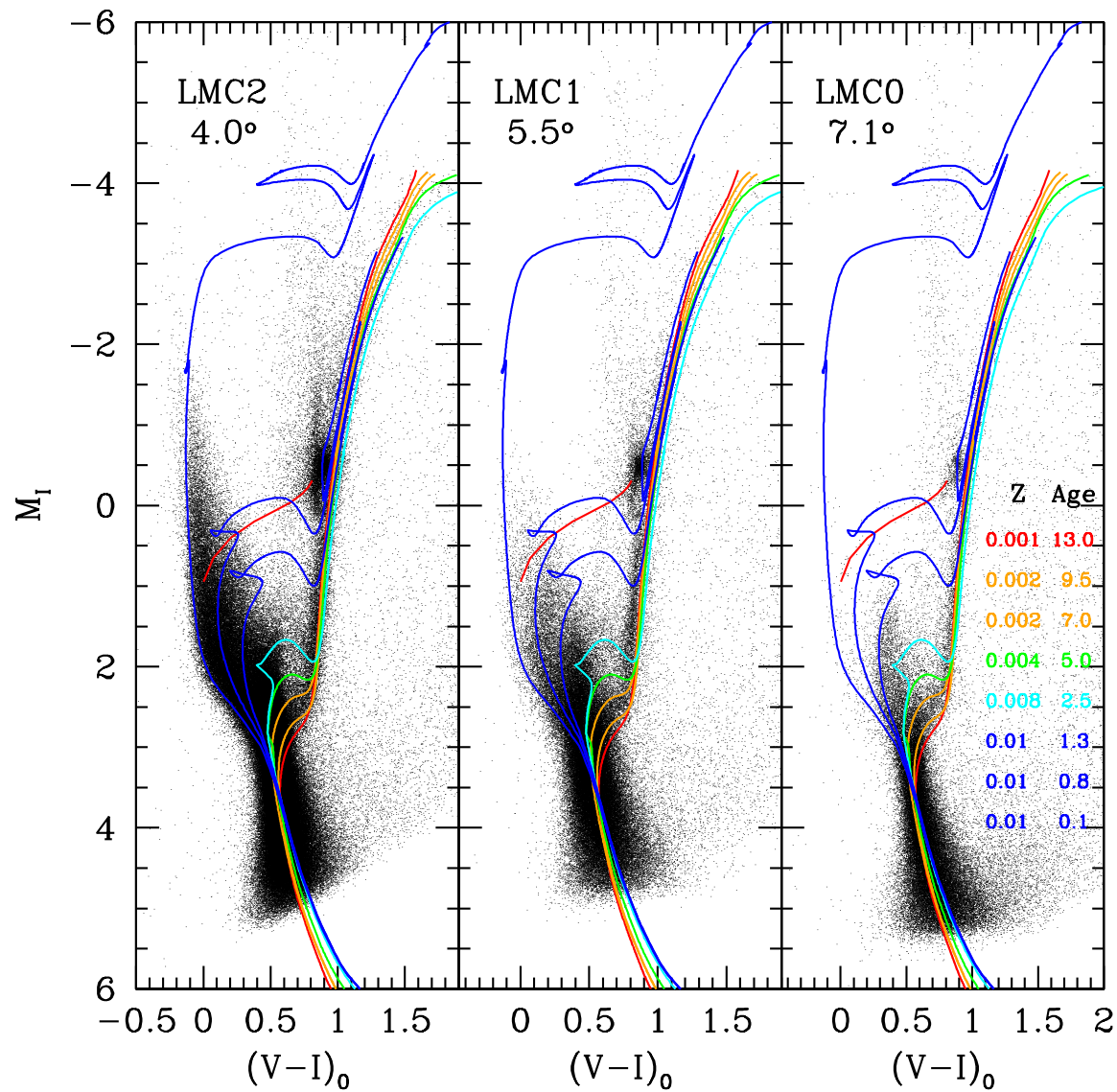}
\caption{[$(V-I)_0$, $M_I$] CMDs for the three fields, as labelled, after applying the quality cuts described in Section~\ref{observations}. The final number of stars are 72000, 114000 and 290000 stars in fields LMC0, LMC1 and LMC2 respectively. Isochrones from \citet{Pietrinferni04} of ages and metallicites as labelled have been superimposed. The three youngest isochrones have been chosen to indicate the adopted age of the end of the bulk of the star formation in fields LMC0 (1.3 Gyr), LMC1 (0.8 Gyr) and LMC2 (0.1 Gyr).}\label{cmds}
\end{figure*}

Figure~\ref{cmds} displays the CMDs of the three LMC fields. BaSTI isochrones \citep{Pietrinferni04} with ages and metallicites as labelled have been superimposed as references for the forthcoming discussion. A distance modulus of (m-M)$_{0}$=18.5  \citep{Freedman01, Alves04, Pietrzyski13} and reddening values as listed in Table~\ref{fields} have been adopted (see Table caption for details). All the CMDs reach a couple of magnitudes below the oMSTO, and therefore, these are cleanly reached with good photometric precision.  Completeness factors have been calculated as the ratio of the number of artificial stars recovered to the number of injected stars in each magnitude interval. At the bright limit, the completeness is approximately 100\%, as expected, in all fields. In the most external field, LMC0, the completeness is over 90\% down to the photometric limit $I \simeq 23$. In the other two fields, it is $\simeq$ 80\% at the level of the oMSTO ($M_I\simeq$3) and 50\% one magnitude fainter.  

Field LMC2, the closest to the LMC dynamical center (R$_{gc}$=4.0\degr\ or 3.5 Kpc), shows a CMD with a prominent, bright main sequence and well populated red-clump (RC), typical from a population which has had a continuous star formation from $\simeq$ 13 Gyr ago to virtually the present time (see below). The two outermost fields, LMC1 and LMC0, (R$_{gc}$=5.5\degr\ and 7.1\degr\ or 4.8 and 6.2 Kpc, respectively, from the LMC center) clearly show a fainter termination of the main sequence, indicative of a truncated or sharply decreasing star formation rate in the last few hundred million years or $\simeq 1$ Gyr, respectively. The comparison with the isochrones displayed in Figure~\ref{cmds} allows us to estimate and age for the end of the main star formation activity in each field \citep[see also]{Gallart08LMC}. These ages are listed in Table~\ref{tabpsi}. Finally, a few red AGB stars can be observed in each CMD, with colours (V-I)$\geq$ 2 (outside the plotted area). No prominent blue horizontal-branch (HB) is observed in any of the fields.


\section{Deriving the Star Formation History} \label{derivation}

The SFHs of the three LMC fields analysed is this paper have been obtained using the suite of codes IAC-star \citep[for synthetic CMD computation;][]{iacstar}, IAC-pop \citep[for solving for the best parameterization of the SFH;][]{iacpop}, and MinnIAC \citep[a suite of routines designed to handle the sampling of the parameter space, creating input data to IAC-pop, and averaging solutions;][]{Hidalgo11sfhlgs3}. The reader is referred to the reference papers for a detailed discussion of the procedures involved. In the following, we will only provide an overview of the process by discussing the particulars of the derivation of the SFH of our LMC fields. 

1. Using IAC-star, we calculated a synthetic CMD with 8$\times10^6$ stars adopting the following inputs: 
 
 i) the BaSTI stellar evolution library \citep[][solar scaled, overshooting set]{Pietrinferni04} with the \citet{Castelli03} bolometric corrections; 

 ii) a constant star formation rate between 13 and $\simeq$ 0.030 Gyr ago (that is, the age limit of the BaSTI library, which varies slightly as a function of the metallicity) and no a-priori age-metallicity relation: stars of all ages are uniformly distributed between Z=0.0001 and Z=0.02. An age of 13 Gyr has been adopted as the LMC formation age taking into account the currently assumed age of the Universe of 13.7 Gyr \citep{Spergel03}, which is also in good agreement with the ages of old Milky Way globular clusters derived by \citet{MarinFranch09} using the same BaSTI stellar evolution models.  The lower metallicity is set by that available in the adopted stellar evolution models, while the upper metallicity limit is slightly larger than the maximum metallicity derived by \citet{Carrera08LMC} for the LMC \cite[see also][]{Carrera11LMC} using CaII triplet spectroscopy of RGB stars, in order to allow the presence in the solution of relatively high metallicity young stars; 

 iii) the \citet{Kroupa07} canonical stellar IMF, namely $N(m)dm = m^{-\alpha} dm$, where $\alpha$=1.3 for stars with mass 0.1$\leq m/M_{\odot} \leq$ 0.5, and $\alpha$=2.3 for 0.5$\leq m/M_{\odot} \leq$ 100. Different values of both exponents have been tested for the galaxies of the LCID project that host current star formation (LGS3 and IC1613) \citep[see]{Hidalgo11sfhlgs3}, and best solutions were obtained with values compatible with the \citet{Kroupa07} IMF; 

 iv) a binary star distribution function $\beta(f,q)$ with a binary fraction $f=0.4$ and relative mass distribution $q > 0.5$. These values are approximately consistent with observations suggesting that a substantial fraction of stars form in binary or multiple systems \citep{Duquennoy91, Lada06}. In addition, \citet{Monelli10sfhcetus} have shown that the adopted parameterization does not have a strong impact on the derived SFH as long as some fraction of binary stars is included. $q > 0.5$ is adopted because only binaries with relatively high mass fractions can be distinguished from single stars in the CMD.  

2. Observational effects, including incompleteness, were simulated in the synthetic CMD using the code $obsersin$ \citep[][see also Gallart et al. 1996b]{Hidalgo11sfhlgs3}. This code uses the information from the completeness tests described in Section~\ref{observations} and recorded in the {\it crowding trial table}. This is a fundamental step because the position of the stars in the observed CMD is substantially affected by the observational effects. We will refer to the synthetic CMD, after the simulation of observational effects,  as model CMD.

\begin{figure}
\centering
\includegraphics[width=8cm]{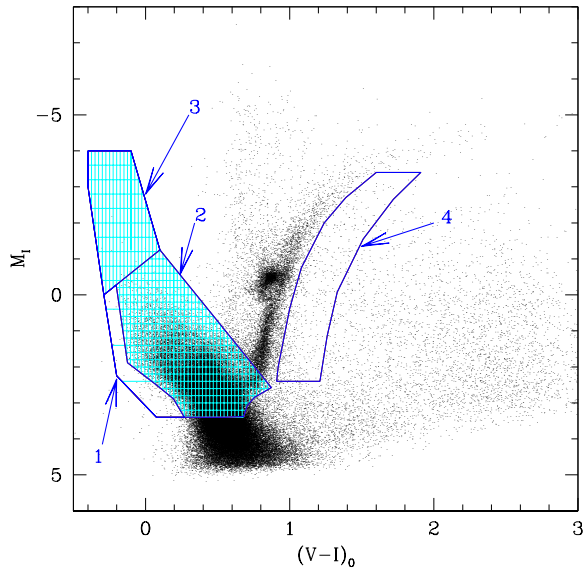}
\caption{Set of bundles and boxes used to count stars in the observed and model CMDs. The sizes of the boxes in each bundle are given in Table~\ref{boxes} }\label{bundles}
\end{figure}

3. The model CMD was sampled by defining (i) the age and metallicity bins that fix the simple populations and (ii) a grid of bundles and boxes in the colour-magnitude plane \citep[see][and Figure~\ref{bundles}]{Hidalgo11sfhlgs3} where the stars in the model and the observed CMD are counted. A distance modulus of (m-M)$_{0}$=18.5 and reddening values as listed in Table~\ref{fields} have been adopted. Differential reddening has been assumed to be negligible in these fields, which are locateed at relatively large galactocentric distances, where the HI column densitites are low (see Section~\ref{discussion}).  This assumption is supported by the tightness of the sequences in the different CMDs, and in particular that of the RC. An small amount of differential  reddening, of the order of hundredths of magnitude, according to the reddening variations within the fields indicated by the \citet{Schlegel98} reddening maps, would 'blur' the stars in the main sequence by an amount that can be easily absorbed by the finite size of the boxes used to sample the CMDs. 

\begin{table}
\centering
\caption{Box sizes in each bundle. Missing quantities indicate values as large as the box dimension itself}
\begin{tabular}{l|ccc|}\\
\hline \hline
Bundle  &$\Delta$mag  &$\Delta$col  &\\
\hline
1   &1.0  &-- \\
\scriptsize  2   &0.2  &0.025 &\\
3   &0.4  &0.025 &\\ 
4   &--  &--    &\\ 
\hline
\label{boxes}
\end{tabular}
\end{table}

The age intervals are of 1 Gyr duration for the first 12 Gyr of the galaxy evolution, and smaller for the last 1 Gyr in order to account for the better age resolution at younger ages. Specifically, the ages that define the limits of the age bins are 0, 0.1, 0.2, 0.5, 1 Gyr and 1-12 Gyr in steps of 1 Gyr. Similarly, the metallicity intervals are wider at higher metallicity. They are, specifically, (0.1, 0.3, 0.5, 0.7, 1, 2,   3, 4, 6, 8, 10, 12, 14, 16, 18, 20) $\times 10^{-3}$. Figure~\ref{bundles} represents the location of the bundles in the observed CMD of field LMC1 and how these bundles are divided in boxes of varying size for each bundle. The sizes of the boxes in each bundle are given in Table~\ref{boxes}. This distribution of bundles closely follows the shape of the main sequence, avoiding areas with important foreground contamination and/or small number of observed stars. The size of the boxes varies according to the lower and higher density of stars populating the main sequence at very young and intermediate ages (bundles 3 and 2). Bundle 1 is useful to constrain the upper metallicity limit of the  young populations, but the boxes are set large in it because the number of stars in that area is small.   Bundle 4 helps to constrain the upper metallicity limit of the intermediate-age and old population. Because of the substantial foreground contamination\footnote{The level of contamination by foreground stars of different areas of the CMD has been estimated using the Besan\c{c}on galaxy model \citep[][]{Robin03}. It has been found that less than 2\% of stars in the main sequence bundles are foreground stars, while most stars in bundle 4 are foreground. We therefore considered the foreground contamination negligible on the main sequence, and estimated to be zero the number of actual LMC stars in bundle 4. Note however, that the weight of this bundle in the final solution is small, though it helps to clean the solution of likely spurious contributions to the solution of simple population of intermediate age and relatively high metallicity. See \citet{Meschin12} for details.} in this bundle, the number of observed stars in it has been forced to be zero, which we believe is an accurate representation of reality. A number of other samplings (sets of  intervals of both age and metallicity, and distributions of bundles and boxes) have been used to derive the SFH in some of the fields. These tests have shown that the derived SFH does not depend strongly on the sampling of the CMDs, since the recovered SFHs are essentially the same within the error bars, independently of the sampling used \citep[see][] {Meschin12}. 

In order to reduce the dependency (even if small) of the resulting SFH on the choice of simple populations and boxes in the CMDs, we calculated the SFH several times, each time shifting the limits of age and metallicity used to define the simple populations, and/or the location of the boxes in which the stars are counted in the CMD. In particular, we shifted the age and metallicity four times, each time by an amount of 25\% of the bin size, with four different configurations: a) moving the age bin toward increasing ages, at fixed metallicity;   b) moving the metallicity bins toward increasing metallicity at fixed age; c) shifting both bins towards increasing values, and d) shifting toward increasing metallicity and decreasing age. With this we obtain 16 positions in the age-metallicity plane (four shifts of the mesh in each of four different configurations). The 16 different samplings were used twice, by shifting the boxes a fraction of their size across the CMD. These 32 solutions were then averaged to obtain a more stable solution. To represent the merit function of this average solution we simply adopt the average of the 32 individual, reduced $\chi^2$, which we denote $\overline{\chi^2}$. 

\begin{figure}
\centering
\includegraphics[width=8cm]{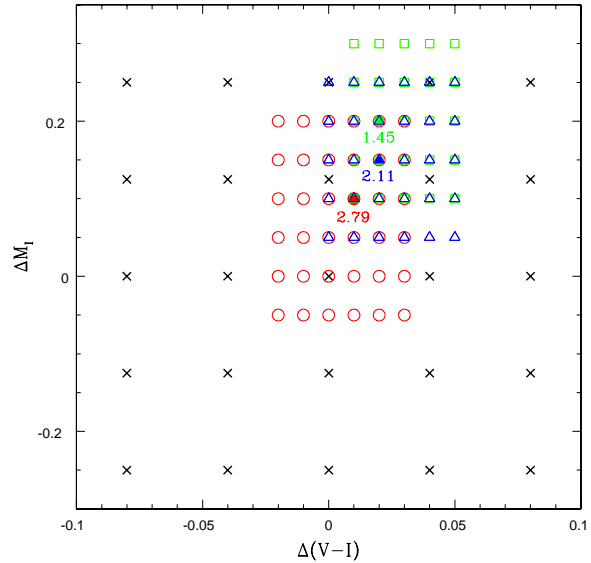}
\caption{Representation of the 5$\times$ 5 grid discussed in the text. Thirty two solutions have been obtained  in each node of the grid, using slightly different simple stellar population definitions and location of the boxes in the CMD. In each node, the 32 solutions are averaged, and the corresponding $\overline{\chi^2}$ is calculated. The nodes of the grid where  $\overline{\chi^2}$ is minimum are indicated, together with its value. } 
\label{mapachi}
\end{figure}

This process is repeated several times after shifting the observed CMD with respect to the model CMD, in order to account for uncertainties in photometric calibration, distance and mean reddening. The CMD was shifted 25 times on a regular 5$\times$5 grid, with nodes $\Delta$(V-I)=[-0.08, -0.04, 0, 0.04, 0.08] and $\Delta I$=[-0.25, -0.15, 0, 0.125, 0.25]. This grid is represented in Figure~\ref{mapachi} (black crosses). In each of the 25 nodes of the grid we calculated the 32 solutions described above, thus obtaining a total of 32 $\times$ 25=800 solutions. In each node, the 32 solutions are averaged and the corresponding $\overline{\chi^2}$ is calculated. In this way, we obtain a ''$\chi^2$ map'', where the minimum $\overline{\chi^2}$, $\overline{\chi^{2}_{min}}$, indicates the position in the grid of the best solution calculated. Once this position is identified, a second series of finer shifts around this minimum were applied to refine its location (those are represented as circles, triangles and squares for fields LMC2, LMC1 and LMC0, respectively). The best solution is obtained in ($\Delta(V-I), \Delta M_I)$=(0.02,0.2) for field LMC0, (0.02,0.15) for LMC1 and (0.01,0.1) for LMC2, with  $\overline{\chi^{2}_{min}}$=1.45, 2.11 and 2.79, respectively. Note that we do not necessarily consider the position of  ${\chi^{2}_{min}}$ in the distance-reddening map as a reliable estimate of distance and reddening, since photometric zero-points, model uncertainties and other hidden systematics may affect its absolute location. However, some of these may be the same for the three fields, and thus, we will check whether, in a differential sense, these shifts agree with the expected distance differences among them.

We have used the latest parameters describing the LMC geometry and center of mass position obtained by \citet[Their Table 1, third column]{VdM13}\footnote{Note that their center of mass position is in good agreement with the kinematic center obtained by \citet{Kim98} from HI data, which we have used in Table~\ref{fields}.} to calculate the $\Delta$ (m-M) expected between the center of mass and each of our fields, and obtained $\Delta$ (m-M)=(-0.21, -0.17,-0.12) for fields LMC0, LMC1 and LMC2, in excellent qualitative and quantitative agreement with the $\Delta M_I$ found in our SFH determination. This reinforces the robustness of our determination of the position of the best fit, and is an additional consistency check of the SFH determination method.


\section{The star formation history of the three LMC fields} \label{sfhs}

\subsection{Solutions and confidence in the solutions} \label{solution}

\begin{figure}
\centering
\includegraphics[width=8cm]{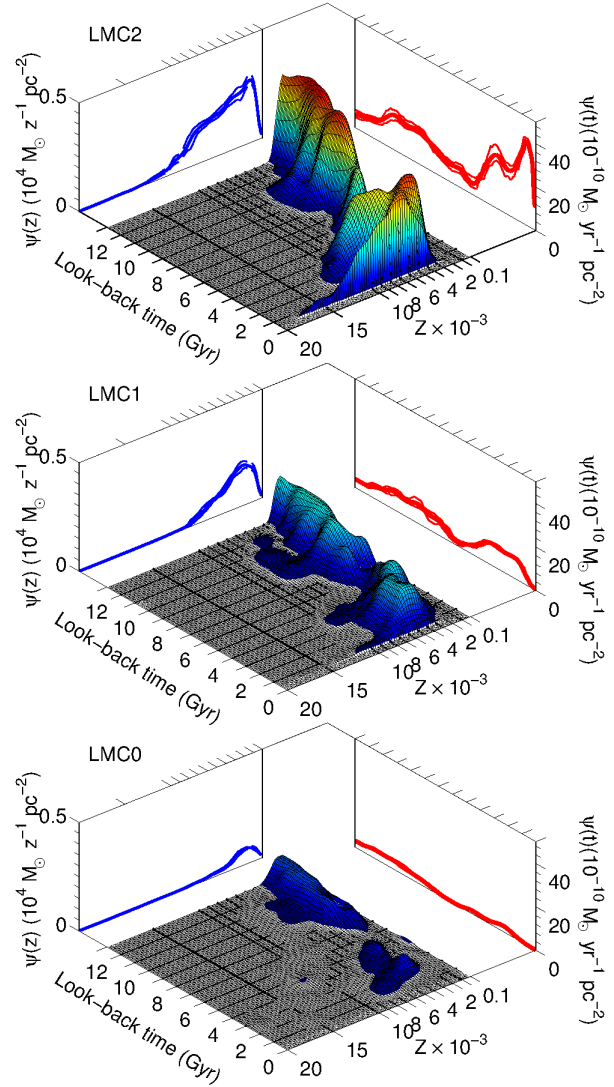}
\caption{Three-dimensional representation of the SFH of the three LMC fields.  The running average of 32 individual solutions at the ${\chi^{2}_{min}}$  position is shown. The colour scale has been kept fixed in the three panels to provide information on the variation of the absolute values of the star formation rate as a function of galactocentric distance. }\label{3sfh3d}
\end{figure}

Figure~\ref{3sfh3d} shows a three-dimensional representation of the SFH of the three LMC fields, $\psi(t,Z)$, as a function of both $t$ and $Z$. $\psi(t)$ and $\psi(Z)$ are also shown in the $\psi-t$ and $\psi-Z$ planes, respectively. The projection of $\psi(t,Z)$ on the age-metallicity plane shows the age-metallicity relation, including the metallicity dispersion as a function of time. These solutions are calculated as the average of the 32 individual solutions calculated at the ${\chi^{2}_{min}}$ in the $\Delta(V-I), \Delta M_I)$ grid \citep[][]{Hidalgo11sfhlgs3}. Note that, thanks to the combinations of shifts in the age-matallicity plane, which define different sets of simple populations with different central values of age and metallicity, it is possible to achieve a higher resolution in the average SFH, as compared to the original size of the age and metallicity bins defining the simple populations. This representation provides an immediate, general visualisation of the SFH of each field, since it includes the information of the star formation rate as a function of both age and metallicity, and the age and metallicity distributions as projections of the first. The colour scale has been kept fixed for the three fields, in order to illustrate the decreasing absolute values of the star formation rate with increasing galactocentric distance.

Error intervals have been represented in the $\psi-t$ plane in Figure~\ref{3sfh3d} (see also Figure~\ref{sfrt}). These are calculated from the dispersion of the 32 solutions  calculated at the ${\chi^{2}_{min}}$ in the $\Delta(V-I), \Delta M_I)$ grid, {\it plus} any solution in the grid that differs in  $\chi^{2}$ by less than 1$\sigma$. This provides errors equal or in excess to the so-called {\it several solutions} criterion which was shown by \citet{iacpop} to produce reliable estimates of total internal errors. 


\begin{figure*}
\includegraphics[width=16cm]{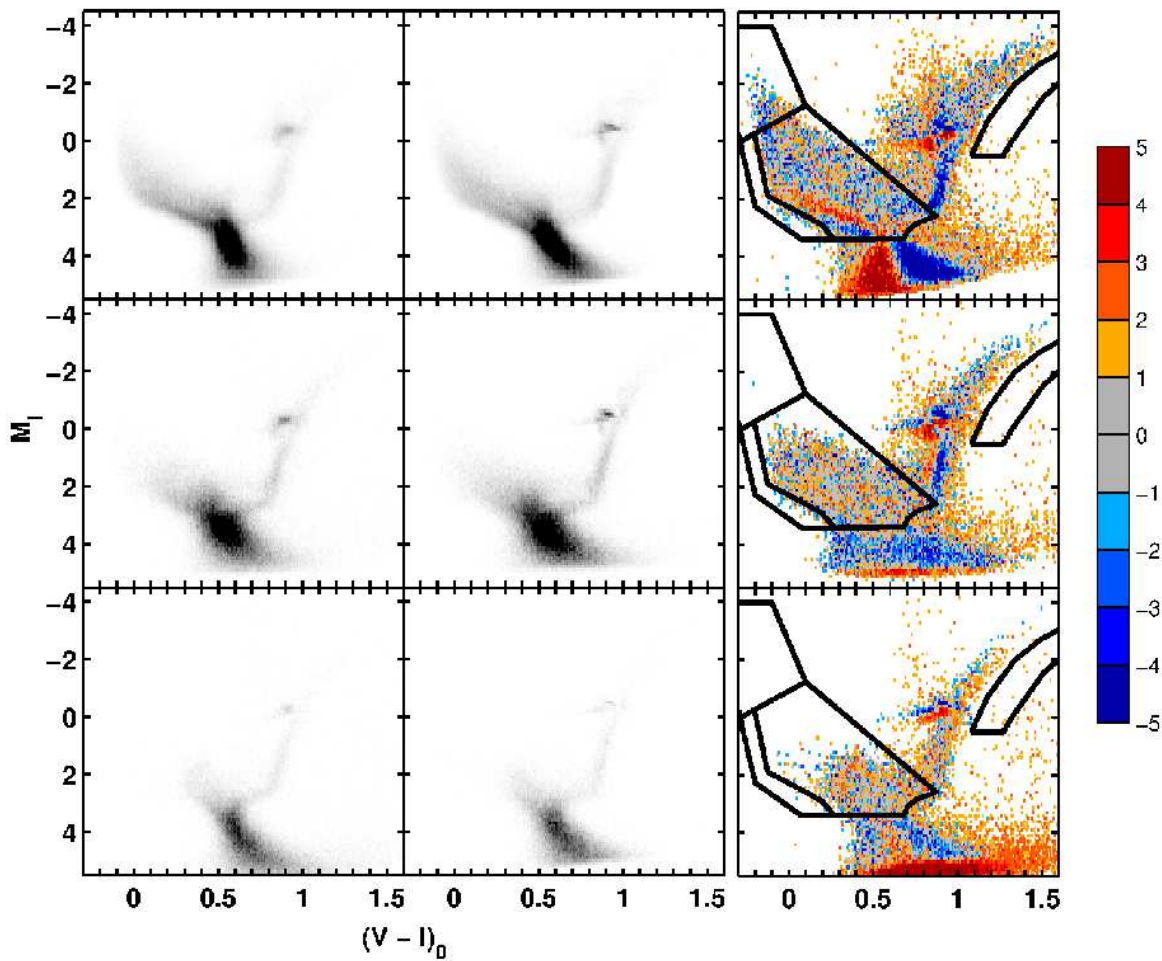} 
\caption{Hess diagrams corresponding to the observed (left panel) and solution (central panel) CMDs and residuals (right panel) color-magnitude diagram for the three LMC fields (LMC2, LMC1 and LMC0 from top to bottom, respectively). The residuals are in units of Poisson uncertainties.}
\label{sol_cmd}
\end{figure*}

Figure~\ref{sol_cmd} displays the Hess diagrams associated to the observed (left panels) and solution CMDs (central panels), for each field (LMC2 to LMC0 from top to bottom). The right panels show the residual diagrams. The bundles that have been used for the SFH determination have been over-plotted in this last panel. An overall good agreement between the observed and the solution CMDs can be observed, particularly in the main-sequence and subgiant-branch areas. Note for example the variation in the stellar density of the subgiant-brach as a function of magnitude,  between $M_I=$ 3 and 1 in fields LMC2 and LMC1, with a lower density area in the middle, visible in both the left and the central panels. In the right panels, the residuals have been expressed in units of Poisson uncertainties. The gray color, which represents the interval $\pm 1 \sigma$ is the predominant one in this panel. Inside de bundles, most cells have $\pm 3 \sigma$ and, most importantly, no signifficant structures can be observed in the residuals map. One possible excepcion to the latter is the narrow stripe with positive sigma between $2<M_I<4$ and $0<(V-I)_0<0.5$ in field LMC2, which seems to be compensated with a smaller area of negative sigma just below it. This feature seems to be the result of the very sharp feature that can be appreciated in the observed Hess diagram near the lower envelope of the main sequence. This feature has not been reproduced as sharply in the Hess diagram corresponding to the solution, possibly because its size in magnitude is similar to the size in magnitude of the boxes in bundle 2, which have therefore, a limited resolving power. Ouside the bundles, relatively high residuals are observed in the RGB portion below the red-clump/HB area, and in the red-clump/HB themselves: the low RGB is slightly too red in the models, while the red-clump shows sharper features in the model than in the observations that produce high residuals. Note hovewer, the good qualitative morphological agreement (e.g. existence of a {\it secondary clump} \citep{Girardi99}  and of a {\it vertical extension of the red clump} \citep[see][]{Gallart98bump}. These mismatches between observed and model CMDs are partially due to the shortcomings of stellar evolution models in these advanced stellar evolutionary phases. It is in fact because of these shortcomings that we routinely don't use these advanced evolutionary phases in the derivation of the SFH \citep{Monelli10sfhcetus}. Finally, the high residuals in the low main sequence, particularly in field LMC2, are due to innaccuracies in the results of the artificial stars tests at this relatively faint magnitude limit, where completeness in below 50\%. In fact, note the aparent different slope of the low main sequence between the observed and model Hess diagram of field LMC2: this discrepancy is hardly noticeable when stars, instead of star densities, are plot which indicates that a minority of stars in the low main sequence are responsible for these large residuals.

\subsection{Features of the SFH of the three LMC fields}

\begin{figure}
\centering
\includegraphics[width=8cm]{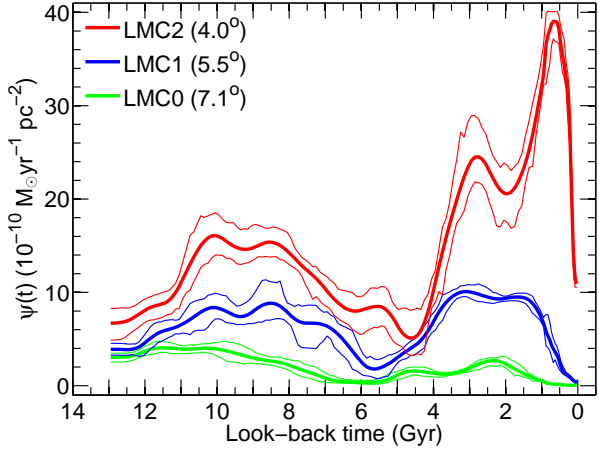}
\caption{Comparison of $psi(t)$ for the three LMC fields. The thin lines represent the uncertainties (see text for details).}\label{sfrt}
\end{figure}

Figure~\ref{sfrt} displays the $\psi(t)$ projection for the three fields. Error intervals calculated as discussed in Section~\ref{solution}, are shown. This figure shows that the galaxy has experienced two main epochs of star formation, separated by a period of lower star forming activity. The first epoch (which we will refer to as {\it old star forming epoch}, ${\rm {O_{SFE}}}$) started at the earliest times we have considered in this study, $\simeq$ 13 Gyr ago, and peaked around 10-8 Gyr ago. The period of low star forming activity extends from $\simeq$ 7  to $\simeq$ 4 Gyr ago, after which a second {\it young star forming epoch} (${\rm {Y_{SFE}}}$ thereafter) occurs, and  lasts till approximately the present time in field LMC2, and till $\simeq$ 0.8 and $\simeq$1.3 Gyr ago in fields LMC1 and LMC0 respectively. The ${\rm {Y_{SFE}}}$ seems to be composed of at least two periods of enhanced star formation activity. Eyeball estimates of their peak ages, together with the approximate age at which the minimum $\psi(t)$ occurs, are listed in Table~\ref{tabpsi}. Note that the age of both peaks gets older at increasing galactocentric distances, as occurs also with the age of the minimum star formation activity.  Table~4 contains a summary of the derived SFH, as a number of values that characterise our solution for the three fields. 

 A very robust feature that can be observed in Figure~\ref{sfrt} (see also Table~4) is that the ratio between the amount of star formation in the ${\rm {Y_{SFE}}}$ and ${\rm {O_{SFE}}}$ (i.e. after and before the minimum $\psi(t)$) decreases with increasing galactocentric radius: ${\rm {Y_{SFE}}}$/${\rm {O_{SFE}}}$= (1.1:0.8:0.4) for (LMC2:LMC1:LMC0). That is, the stellar population is older on average outwards. In the outermost field observed, LMC0, the star formation rate in the second half of the galaxy's life is lower than in the first half, both are similar in field LMC1, while in LMC2, it is substantially higher. In this innermost, field, in addition, the two episodes in which the young star forming epoch can be split have a different star formation rate, the second, younger half being more intense than the first half. These quantitative conclusions are in good agreement with those by \citet{Gallart08LMC} based on comparison of the CMD with isochrones and with observed and theoretical colour functions. They concluded that the colour functions of fields LMC2 and LMC1 suggested a period of enhanced star formation (as compared with a constant star formation rate) in the age range 4-1 Gyr, while the colour function of field LMC2 indicated an additional increase from 1-1.5 Gyr ago to the present time.

\begin{table}
\caption{Eyeball estimates of the age of the end of the bulk of the star formation, as estimated from the comparison with isochrones in Figure~\ref{cmds}, and of the age of various features in $\psi(t)$.}
\label{tabpsi}
\centering
\begin{tabular}{lcccc}\\\hline\hline
Field & End & Peak 1 (Gyr)&Peak 2 (Gyr) &Minimum (Gyr)\\\hline
LMC2  & 0.1$\pm$0.1 &0.7$\pm$0.3 &2.8$\pm$0.3  &4.5$\pm$ 0.5 \\
LMC1  & 0.8$\pm$0.4 &1.5$\pm$ 0.5&3.0$\pm$0.5  &5.5$\pm$ 0.5 \\
LMC0  & 1.3$\pm$0.2 &2.3$\pm$0.5 &4.5$\pm$0.5  &6.0$\pm$ 0.8 \\\hline
\end{tabular}
\end{table}


\begin{figure}
\centering
\includegraphics[width=8cm]{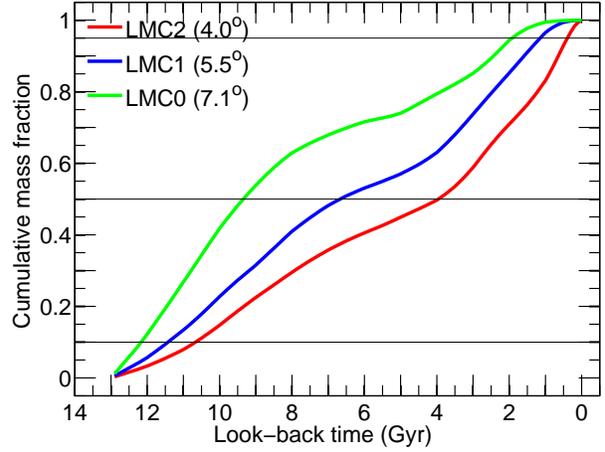}
\caption{Cumulative mass fraction for the three LMC fields. Horizontal lines indicate mass fractions corresponding to 10, 50 and 95\% of the total accumulated mass.}\label{acum}
\end{figure}

These systematic differences in the star formation rate as a function of time in the three fields are reflected in their respective cumulative mass fractions, displayed in Figure~\ref{acum}. In this figure, three horizontal lines have been drawn to indicate the 10\%, 50\% and 95\% mass percentiles, and the ages ($T_{10\%},T_{50\%}, T_{95\%}$) at which these percentiles are reached in each field (see also Table~4). The trend of increasing age at increasing galactocentric distance holds for all three age indicators, which describe a {\it faster} evolution in the external parts of the galaxy. 
All three fields have formed 10\% of their stellar mass before $\simeq$ 11 Gyr ago. $T_{50\%}$, the age at which half of the stellar mass was formed in each field, is 4.0, 6.8 and 9.3 Gyr for LMC2, LMC1 and LMC0, reflecting the substantially different evolution of the three fields. Finally, $T_{95\%}$ indicates the age at which star formation has {\it almost} ended. It is 0.4, 1.2 and 2.0 Gyr from the inner to the outermost field. These are good upper limits of the estimates of the time at which the main star formation activity has ended in each field (first column in Table~\ref{tabpsi}, as discussed in Section~\ref{sec_cmds}, and agree with the conclusions by \citet{Gallart08LMC} based on isochrones and colour functions.

\begin{table}
\label{summ_sfh}
\caption{Derived values from the SFH.}
\begin{tabular}{lccc}\\
\hline \hline
 &LMC2  &LMC1  &LMC0\\
\hline
$\int_{\rm\,O_{SFE}} \psi(t)\,dt \, [10^6 M_{\odot}$] 	       			&2.40$\pm$0.49   &1.22$\pm$0.29   &0.50$\pm$0.14\\
$\int_{\rm\,Y_{SFE}} \psi(t)\,dt \, [10^6 M_{\odot}$] 	 			&2.65$\pm$0.25   &0.96$\pm$0.24   &0.20$\pm$0.05\\
$\int_{\rm \,Y_{SFE}}  \psi(t)\,dt$  / $\int_{\rm \,O_{SFE}}  \psi(t)\,dt$   	&1.10            &0.79            &0.40\\
$\int \psi(t)\,dt \, [10^6 M_{\odot}$]		                   		&5.05$\pm$0.75	 &2.18$\pm$0.41   &0.70$\pm$0.19 \\
$T_{\,10 \%}  \,\,$ [Gyr]					    			&10.8	      &11.5	       &12.2\\
$T_{\,50 \%}  \,\,$ [Gyr]					    			&4.0		&6.8	       &9.3\\
$T_{\,95 \%}  \,\,$ [Gyr]					    			&0.4		&1.2	       &2.0\\
\hline
\end{tabular}
\end{table}


\subsection{The age-metallicity relation}

\begin{figure}
\centering
\includegraphics[width=8cm]{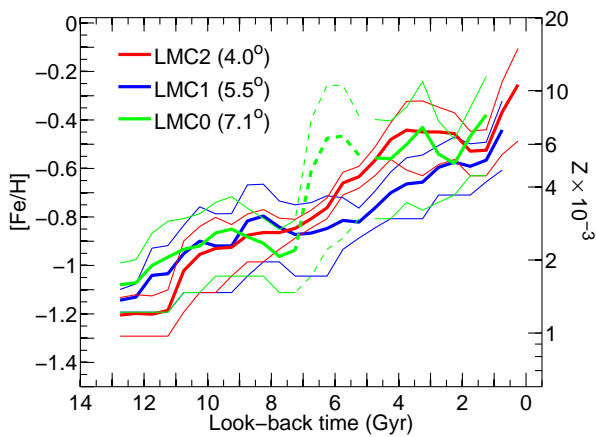}
\caption{Age-metallicity relation, Z(t), for the three LMC fields.}\label{zt}
\end{figure}

Figure~\ref{zt} displays the age-metallicity relation, Z(t), for the three fields. Thick lines represent the mean metallicity at a given age, while thin lines indicate the metallicity interval including 68\% of the metallicity distribution. The figure shows a clear trend of increasing metallicity with time in all three fields, and that the mean metallicity as a function of time is consistent
with a single value for all of them \citep[in good agreement with the conclusions by][]{Carrera08LMC}. We infer an initial mean [Fe/H]$\simeq$-1.1 dex and a metallicity of [Fe/H] $\simeq$ -0.4 dex one Gyr ago. The relatively high initial mean value of [Fe/H] inferred is probably indicating a fast initial chemical enrichment which we cannot resolve in our derived SFH. In the case of field LMC0, the part of the age-metallicity relation corresponding to a look-back time of $\simeq$7-5 Gyr ago has been drawn with a dashed line to highlight that it is subject to important uncertainties, related to the very low star formation activity in this field at this age. However, the fact that we infer an important metal enrichment in this period of little star forming activity is a feature that has also been observed in a number of dwarf galaxies, particularly those of the LCID project (Hidalgo 2014, in prep).

\begin{figure}
\centering
\includegraphics[width=8cm]{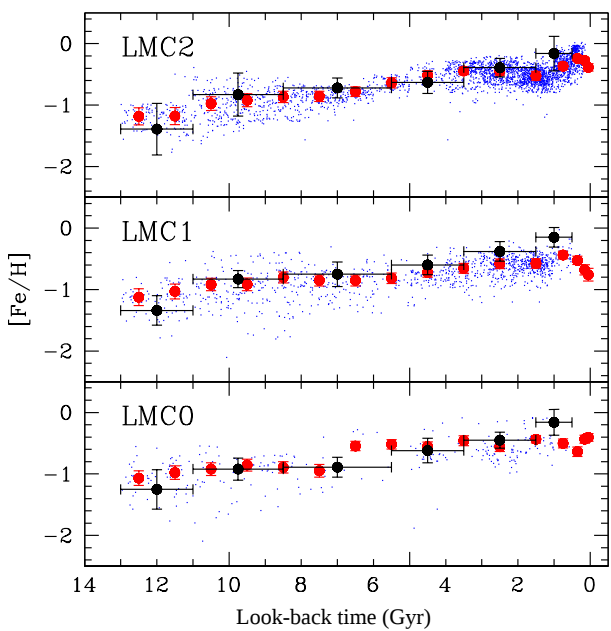}
\caption{Comparison between the age-metallicity relation obtained in this paper using basically information from stars on the main sequence to derive the SFH --red-- and that derived by \citet{Carrera08LMC} from RGB Ca II triplet spectroscopy and photometry --black--. For the \citet{Carrera08LMC} values, the average metallicity and its dispersion have been represented for each age range (horizontal bars). The blue dots indicate age and metallicity of RGB stars in our solution CMD  which lie in the CMD areas used by \citet{Carrera08LMC} to select spectroscopic candidates.}
\label{zt_carrera}
\end{figure}

The age-metallicity relations derived solely from the CMD (and basically using the information on the main sequence) can be compared with those derived by \citet{Carrera08LMC} using RGB Ca II triplet spectroscopy and photometry of RGB stars. In Figure~\ref{zt_carrera} we have represented the average values of [Fe/H] for six age bins (black circles), as derived by \citet{Carrera08LMC}, together with mean values of metallicity as derived from our SFH (red circles) at each age interval considered, and ages and metallicities of RGB stars in our solution CMD, which lie in the CMD areas used by \citet{Carrera08LMC} to select spectroscopic candidates. Note the excellent agreement of the three sets of indicators of Z(t) in the range of validity of the spectroscopic determinations (age $>$ 1 Gyr). Only one point for the LMC0 field, at age $\simeq 6.5$ Gyr shows some disagreement between both sets of determinations. This discrepancy may be due to the little statistics of stars in this age range,  which coincides with the minimum star formation rate in the less populated field, LMC0. This result is another important external consistency demonstration of our procedure to determine the SFH. 


\section{Comparison with previous results.} \label{prev}

\subsection{Comparison with field SFHs.} \label{compara_sfh}

The published LMC field SFHs obtained from CMDs reaching the oldest MSTO with high photometric precision are based on WFPC2 observations. In the case of bar fields, the CMDs are relatively well populated  \citep[e.g.][]{Holtzman99, Olsen99, Smecker-Hane02,Weisz13}, and yield relatively consistent SFHs, with star formation rate depressed from $\simeq$ 10 to 6 Gyr ago  \citep[see][for a comparison of three different determinations for old ages]{HarrisZaritsky09} and somewhat increased or bursty in the last couple Gyr.  Due to the small field of view, the WFPC2 disk CMDs are sparsely populated down to the oMSTO \citep[][]{Holtzman99, Javiel05}. \citet{Smecker-Hane02} tried to overcome this problem by mosaicking 10 WFPC2 fields in a disk field located $\simeq$ 1.7\degr\ (1.5 Kpc) Southwest of the LMC center. While the corresponding SFHs present field-to-field variations, particularly at the young side, the general trend is, in contrast to the cluster age distribution, a relatively flat star formation rate as a function of time, with mild enhancements in the last few Gyr, but star formation not always continuing to the present time.  While the variations at young ages may be related to the actual gradient in the stellar populations present across the LMC disk, it is quite possible that the scatter in the field SFH results is related to small number statistics in the CMDs (and to the different analysis techniques among authors). In fact, our results based on well populated CMDs show  features common to all fields and consistent trends as a function of galactocentric distance. A detailed comparison with these results, therefore, does not seem appropriate.

Two ground based projects are providing wide-field coverage of both Magellanic Clouds (MC): the Magellanic Clouds Photometric Survey \citep[MCPS:][]{Zaritsky97MCPS} and the Vista Magellanic Clouds Survey \citep[VMC:][]{Cioni11}. 

\begin{figure}
\centering
  \includegraphics[width=8cm]{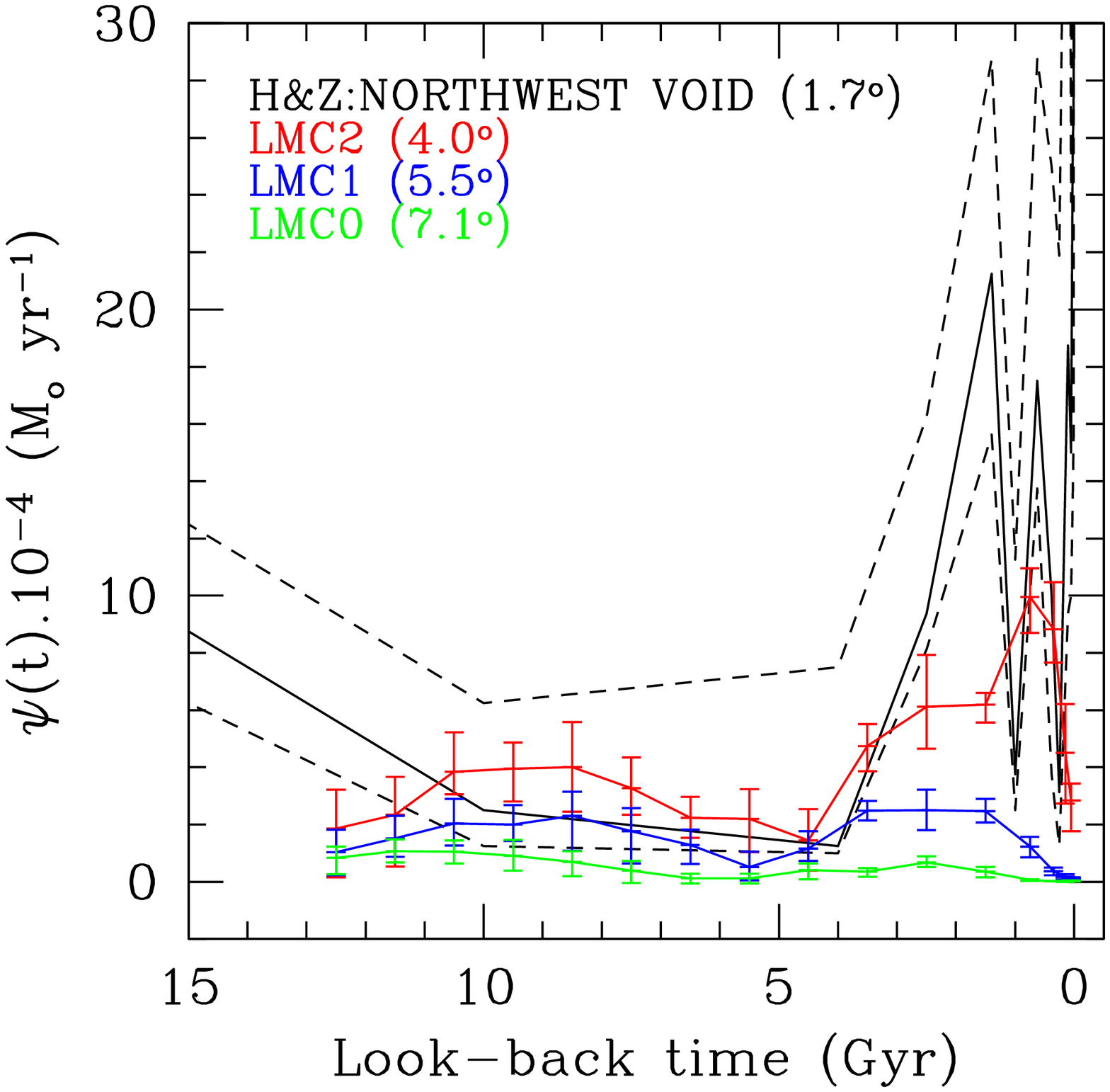}
  \includegraphics[width=8cm]{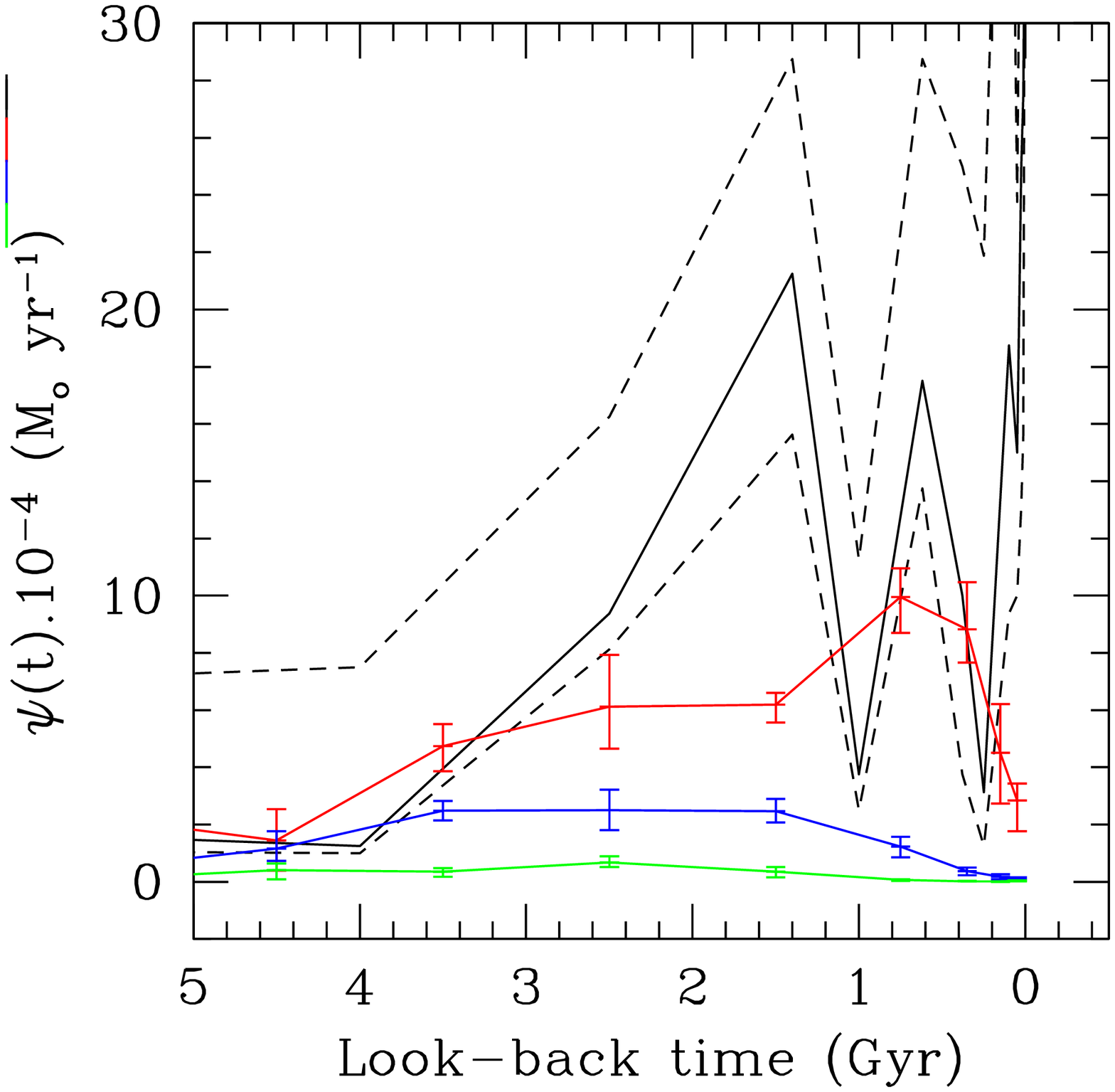}
\caption{Comparison between the SFH obtained by \citet{HarrisZaritsky09} for the region they refer as Northwest Void and the SFHs of our fields. The first has been scaled to the area of our fields. Dashed lines in the \citet{HarrisZaritsky09} solution indicate their quoted uncertainties. The lower panel displays a close-up view of the last 5 Gyr.  Note the good morphological agreement, within the errors, between the SFH of the Northwest void and that of our field LMC2. The former is 2.5\degr\ closer to the center than field LMC2. }
\label{nwv}
\end{figure}

\citet{HarrisZaritsky09} obtained a spatially resolved SFH for the LMC covering the central 8.5\degr$\times$7.5\degr. They anchored the SFH at ages older than 4 Gyr to a 'consensus' SFH based on the HST results of \citet{Olsen99}, \citet{Holtzman99} and \citet{Smecker-Hane02}, and thus, their resulting SFH retains the common feature mentioned above of a quiescent star formation epoch from approximately 12 to 5 Gyr ago. After that, they found that star formation proceeded until the current time at an average rate with temporal variations at a factor of  2 level. They find peaks at roughly 2 Gyr, 500 Myr, 100 Myr and 12 Myr. In addition to this global SFH, \citet{HarrisZaritsky09} also discuss the SFH of several localised features in the LMC. The field that they call Northwest Void, which is located in the same approximate direction as our fields, is among them. Its derived SFH is found similar to that of the LMC as a whole, with no local recent deviations from the average behaviour. Our fields have also been chosen not to present particular features that could be related to recent star formation enhancements and this, together with the relative spatial localisation, indicates that a comparison between the SFH results in our fields and those in the Northwest Void may be meaningful. Figure ~\ref{nwv} shows the comparison between the SFH for the Northwest Void  and that of our fields. There is a good overall agreement between the SFHs, with the main features in our solution also present in that of the Void, and the galactocentric trends also extending to the SFH of the Void.   For ages $\leq$ 4 Gyr, for which the SFH of the Void is actually calculated from the data, the star formation rate enhancement relative to previous epochs is present, and more prominent in the void than in our fields, thus continuing with the galactocentric trend we found. Note that the Void SFH solution shows four peaks located at $\simeq$ 1400, 600, 100  and 20 Myr. If we identify the two oldest ones with the two we find in our solutions, they continue the trend of being located at younger ages for decreasing galactocentric distance. The youngest peak continues the trend of continuing star formation to younger ages inwards.

\citet{Rubele12} analysed three VMC tiles located $\simeq$3.5\degr (3 Kpc) away from the LMC center in different directions. Our field LMC2 is located 1.5\degr\ Northwest from the center of tile 8$\_$3. They derive the SFH for a number of 21.0\arcmin$\times$21.5\arcmin sub-regions in which the tiles are subdivided and find that i) the young SFH varies substantially from field to field, and ii) the presence of two peaks in the SFR$(t)$ at ages log($t/$yr) $\simeq$9.3 and 9.7 (approximately 2 and 5 Gyr ago) in most sub-regions, and in particular in those of tile 8$\_$3. While the peak 2 Gyr ago might be identified with our young peak with a similar age, the one found 5 Gyr ago would coincide right in the middle of our epoch of lowest star formation activity. We conclude, therefore, that our SFH is not in agreement with those derived (with substantially lower age resolution) by \citet{Rubele12} from less deep VMC data.  

\subsection{Comparison with clusters star formation and chemical enrichment history} \label{comparison_clusters}

The LMC cluster system has been extensively studied. A lot of effort has been invested in characterising its basic properties (e.g. mass, age, metallicity). With this knowledge, it is possible to determine the age-metallicity relation of the cluster population, and the age distribution. A number of clusters with ages as old as those of the old Milky Way globular clusters exist in the LMC \citep{Johnson99, Olsen99, Johnson06}. Then, there is a dearth of clusters formed between 10 and 3 Gyr ago (the so-called age-gap). Only one cluster, ESO 121-SC03, has an age within the age-gap, and some authors \citep[e.g.,][]{Bica98,BekkiChiba07} suggest that it may have been captured from the SMC. Subsequently, and to the present time, a large population of clusters, a fraction of them very massive, as the Milky Way globular clusters, have formed in the LMC.  In this section we will perform a comparison between some characteristics of the cluster population and the corresponding ones for the field.

\begin{figure}
\centering
\includegraphics[width=8cm]{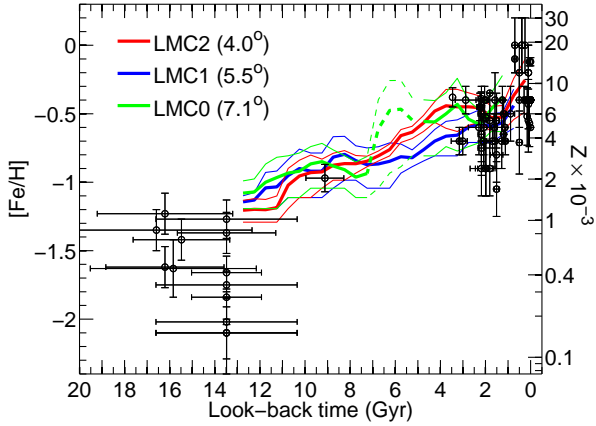}
\caption{Age-metallicity relation of star clusters obtained from the literature (black circles), superimposed to the field star age-metallicity relation derived in this work. In each cluster symbol, errors in age and metallicity have been indicated.}
\label{clus_age_z}
\end {figure}

Figure \ref{clus_age_z} shows the field stars age-metallicity relation derived in this paper for each of the three fields, together with the ages and metallicities of 102 clusters from the literature \citep{Olszewski91, Dirsch00, Bica08, HarrisZaritsky09, Balbinot10}.

A very evident feature that can be observed in this figure is the age-gap, which implies that it is not possible to derive a complete age-metallicity relation from the cluster population. The information that can be extracted from clusters is that, somehow, metallicity increased from [Fe/H] $\simeq$-1.2 for the more metal rich old clusters, to [Fe/H] $\simeq$ -0.7$-$ -0.5 for the clusters catalogued just after the age-gap. At early times, the LMC had formed a number of relatively metal poor, old globular clusters, in a range of metallicity (the range of ages, with some ages older than the currently adopted age of the Universe, correspond to age determinations with old stellar evolution models and/or other not up-to-date adopted parameters for the clusters, e.g. distance).

The correspondence between the cluster age-metallicity relation and that derived for field stars is excellent, with the field age-metallicity relation nicely linking the metallicity of the most metal-rich old clusters and that of the first clusters formed after the age gap. Cluster ESO121-SC03 has a metallicity very much consistent with the metallicity of the field stars of the same age. Interestingly, the metallicity increase at age $\simeq0.5$ Gyr in the two innermost fields is also observed in the clusters. Overall, the intermediate-age and young clusters show a good agreement with the field age-metallicity relation, though a number of clusters of ages 1--2 Gyr have lower metallicity than that of field stars. It would be interesting to investigate whether some of these lower metallicity clusters might have their origin in the SMC, as seems to be the case for a small fraction of LMC stars with origin in the SMC \citep{Olsen11}.  Finally, the substantial dispersion in metallicity of the old clusters is in agreement with the idea discussed in Section~\ref{sfhs} regarding a very fast metal enrichment at the early stages of the LMC evolution.

The age-gap observed in the cluster population is not strictly observed in the field SFH, which shows some level of star formation at all ages. However, the epoch of lower star formation activity that can be observed in all three fields, between $\simeq 8-4$  Gyr ago, qualitatively coincides with the cluster age-gap. It seems that in this epoch of less active star formation, the conditions in the galaxy were not suitable for cluster formation, or that the clusters formed in this period were not massive enough to survive till the present time.  A more comprehensive characterization of the SFH in this epoch across the LMC disk, including more internal regions that contain an important fraction of the mass, will be important to set constraints on the physical conditions for cluster formation.

\section{Discussion} \label{discussion}

\subsection{Stellar population gradients as tracers of galaxy formation and evolution} \label{gradients}

\begin{figure}
\centering
\includegraphics[width=8cm]{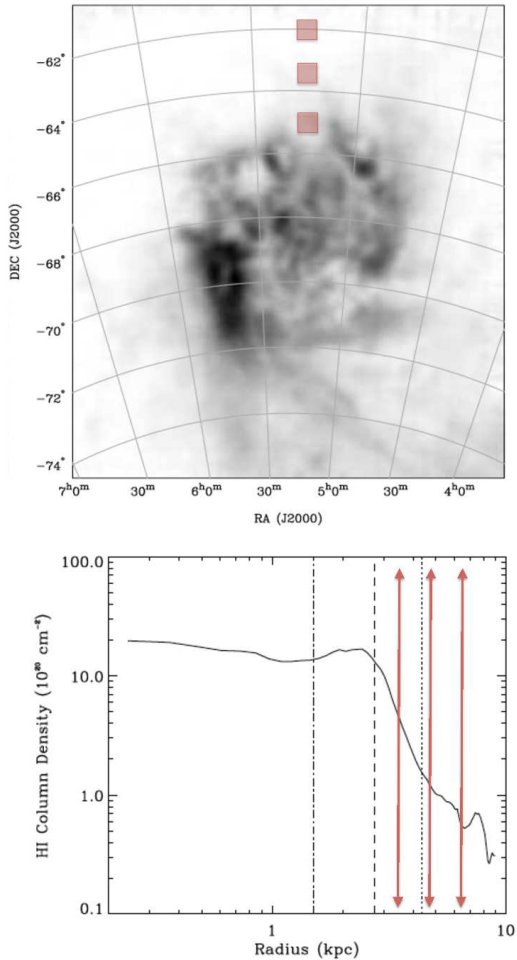}
\caption{Positions of our fields in relation to the HI gas. \citep[HI distributions from][]{Staveley-Smith03}}
\label{hi}
\end {figure}

The SFHs presented in this paper show, in agreement with the analysis presented in \citet{Gallart08LMC} based in the comparison with isochrones and colour functions, that the age of the bulk of the youngest stellar population  gradually increases toward larger galactocentric distances, from currently active star formation in the innermost field, to $\simeq$ 0.8  Gyr and $\simeq$1.3 Gyr in the fields at  5.5\degr and 7.1\degr (4.8 and 6.2 Kpc), respectively. Our SFHs demonstrate also that,  in the $\simeq$ second half of the galaxy's life, the star formation activity has been more intense, as compared with the first half, in the fields closer to the center.

Figure~\ref{hi} displays the position of our three fields, in relation to the HI distribution. The age  gradient of the youngest population is correlated with the HI column density as measured by \citet{Staveley-Smith03}. The innermost field is located $\simeq$ 0.7 kpc beyond R$_{H_\alpha}$ \citep{Kim99}, at a position where the azimuthally averaged and local HI column densities are only $\simeq 5 \times 10^{20} \rm{cm}^{-2}$ and $\simeq 1.8 \times 10^{20} \rm{cm}^{-2}$, respectively. This HI content is somewhat lower than the HI threshold for star formation suggested by \citet{Kennicutt89} or \citet{Skillman87}, but consistent with the fact that lower HI thresholds have been observed in a number of cases \citep[e.g.,][and references therein]{Noordermeer05}. Note, however, that, as discussed in \citet[see also Figure~\ref{cmds}]{Gallart08LMC}, even though there are stars in the CMD of field LMC2 that match a 30 Myr isochrone, a clear step in the density of stars is observed at the position of the 100 Myr isochrone, indicating that the bulk of stars  in this field are older than 100 Myr.  The two outermost fields are close to the HI radius considered by  \citet{Staveley-Smith03} and have a measured HI column density of $3.4\times10^{19} \rm{cm}^{-2}$ and $1.3 \times 10^{18} \rm{cm}^{-2}$, clearly below the HI star formation threshold. Indeed in these fields no current star formation is observed. 


Under the reasonable assumption of no substantial stellar migration across the LMC disk\footnote{In large disk galaxies, there is a consensus both from the theoretical and the observational side, that radial stellar migration caused by gravitational interactions between stars and large scale spiral structures, has a strong effect in the distribution of stellar ages and metallicities as a function of radius \citep[e.g.,][]{SellwoodBinney02, Roskar08, Roskar12, DiMatteoHaywood13, Haywood08}.   At the opposite extreme, in dwarf galaxies simulations show that a very limited amount of radial stellar migration is expected \citep[e.g.,][]{Stinson09, Schroyen13}. To our knowledge, no simulations exist for a galaxy of the LMC characteristics, but the LMC is more similar to a dwarf than to a large spiral in the sense that it does not have a strong spiral structure. We believe, therefore, that the assumption of little radial migration for the LMC is reasonable.}, and if the amount of star formation is related  to the amount of available gas \citep{Kennicutt89}, it can be then inferred that the gas disk  (i.e. with HI density close to the star formation threshold) was substantially larger 1 Gyr ago, when it may  have reached the position of field LMC0 (R$\simeq$7.1\degr or 6.2 Kpc). Note that, for the LMC, the ratio of the HI diameter to the blue diameter, $D_{HI}/D_{25}$ = 1.1 \citep{Staveley-Smith03}, is substantially smaller than the typical value found by \citet{Wilcots96} for a sample of barred magellanic spirals, namely $D_{HI}/D_{25} \geq$ 1.7. A substantially more extended HI disk 1 Gyr ago (even with a corresponding increase of $D_{25}$) would likely make the LMC value closer to those of other SBm galaxies in the field.

This would imply, therefore,  that  the LMC gas disk has "shrunk"  from a R$_{gc}$ of 7\degr\  to 4\degr\ (or from 6.2 to 3.5 Kpc) in 1 Gyr (i.e., at a rate of $\simeq$ 2.7 Kpc/Gyr). This radial shrinking of the gas disk, leading to star formation that is systematically less radially extended with time is qualitatively observed in the simulations of  dwarf galaxies by  \citet{Stinson09}. In these simulations, while stellar migration naturally takes place, the radial age gradient predicted for the stellar populations is stablished at formation rather than due to stellar migration. This occurs as the envelope that contains star formation contracts as the supply of gas diminishes and the pressure support is reduced. 

It may also be that some external process has swiped up part of the HI gas in the LMC \citep[e.g.][has shown that ram-pressure from a low-density ionized Milky Way halo may remove a substantial amount of HI gas from the LMC disk]{Mastropietro05}, or compressed the gas in an azimuthally asymmetric way \citep[such as predicted by][as a result of ram-pressure from the Milky Way hot halo acting on the LMC's interstellar medium]{Mastropietro09}. The timing of either process, occurring in the last $\simeq$ 1 Gyr, as indicated by the age of the last star forming event at the different galactocentric radius, may coincide with the first time that the LMC entered the Milky Way's virial radius \citep[e.g. as in one of the favoured models by][see their Figure 13; see also Salem, Besla \& Bryan 2013, in prep]{Kallivayalil13}. Some of the \citet{Mastropietro09} models including star formation  predict an outside-in migration of the star formation in the outer LMC disk, which specific characteristics depend on the particular assumptions of the model, but are in qualitative agreement with the observations.

\begin{figure}
\centering
\includegraphics[width=8cm]{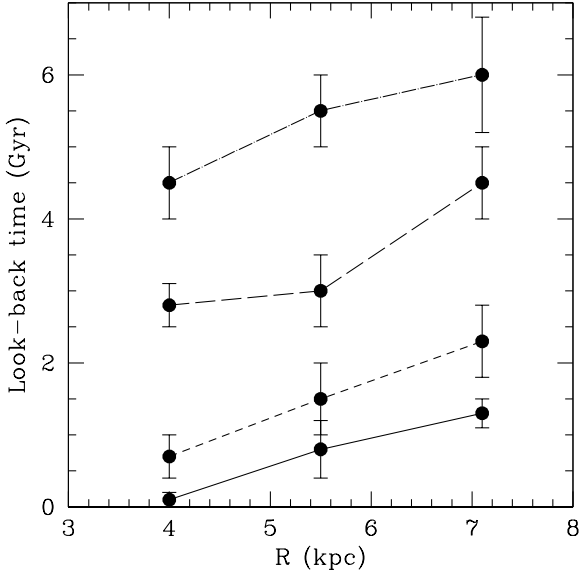}
\caption{Look-back time at which various maxima and minima of star formation are observed in our solutions, as a function of the galactocentric radius of the field (see Table~4).}\label{figpsi}
\end{figure}

Note however, that there are hints in our solutions of a systematic displacement of the ages of the maximum and minimum star formation activity as a function of radius in the second half of the galaxy's history, as quantified in Table~\ref{tabpsi} and illustrated in Figure~\ref{figpsi}. In this Figure, it can be seen that various features in the SFH migrate as a function of radius at a similar rate. However, a total cessation of the star formation didn't happen in any of these fields in the past. 


It might also be that stellar migration plays a role in shaping the current distribution of stellar ages across the LMC disk. The increase of the mean stellar age beyond a profile break, which is hinted at R$_{gc}\simeq$3\degr\ in Figure 1 of \citet{Gallart04LMC} and in Figure 3 of \citet{vanderMarel01}, is an expected signature of radial migration in large disk galaxies forming inside-out. In the case of the LMC, the picture may be further complicated by the presence of a bar, and by its (still uncertain) interaction history. Specific simulations of a LMC like galaxy, ideally taking into account all these factors would be important to assess the predicted importance of stellar radial migration versus in-situ star formation in shaping the LMC stellar population gradients. The possibility of comparing  with detailed SFH profiles, which we aim to obtain for both a larger sample of galactocentric radius and azimuths, makes of the LMC an important test bed of our understanding of the involved physical processes.

\subsection{LMC Orbit and SFH} \label{orbits_sfh}

Under the hypothesis that galaxy interactions trigger star formation, it is common that a determination of the SFH of the Magellanic Clouds is followed by a discussion on the correlations between orbital pericentric passages and maxima of star formation activity \citep[e.g.][]{HarrisZaritsky04}. Likewise, it is common that models  on the formation and evolution of the Magellanic System check their predictions against features of published SFHs \citep[e.g.][]{DiazBekki12}. However, it is quite possible that this is a premature exercise.  On the one hand, in spite of the high accuracy of the proper motions measured using a third epoch HST data \citep{Kallivayalil13}, the orbits of the MCs are still quite uncertain. Remaining significant unknowns  are the structure of the Clouds and their distance, which limit the accuracy of the calculated transverse velocities, and the LMC and the Milky Way masses, which sensibly affect the orbit calculation. On the other hand, the SFH determinations don't converge to provide a coherent view of the evolutionary history of the MCs. We believe this is mostly due to limitations of the available data,  either because the CMDs used are too shallow for a detailed and accurate determination of the SFH {\it at old and intermediate ages} \citep[e.g.][]{HarrisZaritsky04,HarrisZaritsky09, Rubele12} or because they still sample areas too small to provide a representative, global SFH (as is the case of the HST data, see Section~\ref{prev}). The present work presents an intermediate situation, with CMDs reaching the oMSTO and containing  a large number of stars  that ensure a robust SFH determination. The fact that the SFH shows coherent features in the different fields adds some additional confidence to the results.

The LMC HST proper motions derived by \citet{Kallivayalil06LMC} and \citet{Piatek08}, and refined in \citet{Kallivayalil13} have led to a new orbital scenario in which the LMC may no longer be an all-time, bound Milky Way satellite, but it may be approaching our Galaxy for the first time \citep{Besla07}. Even though solutions in which the LMC may have a very long-period, eccentric orbit, and be currently experiencing its second pericentric passage about the Milky Way \citep{ShattowLoeb09, Kallivayalil13} are not totally ruled out given the uncertainties, mainly on the Milky Way and LMC masses,  a number of arguments make a previous pericentric passage unlikely.  These include expectations on the orbital eccentricity coming from cosmological simulations \citep{BoylanKolchin11}, and a number of observational evidence, such as the large gas content of both MCs and the existence of stellar populations extending out to $\simeq$20 Kpc from the LMC's dynamical center \citep{Munoz06, Majewski09IAU, Saha10}. In this context, the classical models of the formation of the Magellanic Stream that relied in repeated close passages of the LMC about the Milky Way have been superseded by models that invoke tidal effects of the LMC on the SMC, under different assumptions for the mutual orbital history. It is now generally accepted that it is sufficient that both MCs have come together a few Gyr ago to satisfactorily explain, for example, the observations of the common HI envelope \citep{Bekki08, Ruzicka10}, the bridge and the Magellanic stream \citep{Nidever08, Besla10, DiazBekki11},  or the {\it age gap} \citep{DiazBekki11}.

In this context, and particularly given the impossibility of proposing a unique orbital scenario \citep{Kallivayalil13}, it does not seem timely to look for detailed coincidences of SFH and orbital features.  What may be found, however, is a qualitative agreement between the common features of the Magellanic System emerging models, derived from the new proper motion measurements and remaining considerations, and the broad features of the LMC SFH. The {\it slow} SFH of the LMC previous to the epoch of depressed star formation found in all fields around $\simeq$6 Gyr ago (compared to  the subsequent, more active periods in the two innermost fields studied) may be associated to a long epoch of evolution of this galaxy in relative isolation before any interaction with the Milky Way. The second part of the LMC's history is characterized by a clearly enhanced star formation activity, since $\simeq 4$ Gyr ago.  The onset of this period may coincide with the LMC approaching the Milky Way, beginning its interaction with the SMC \citep[though][find that the LMC SFH is not very much affected by the interaction with the SMC]{Besla12}, or both. Note that the epoch of depressed star formation between 6-4 Gyr ago would approximately coincide with a previous pericentric passage for the shortest possible orbital periods ($\simeq$ 4 Gyr) obtained by \citet{Kallivayalil13} (see their Figure 11). Under the assumption that such a close passage would have triggered an episode of star formation in the LMC, we think our derived SFHs may be a further argument against this scenario. In contrast, a recent (first) approach to the Milky Way about 1 Gyr ago could be responsible, through  ram pressure effects on the LMC interstellar medium, of a reshaping of the HI profile and the outside-in pattern of the last star forming events (see section~\ref{gradients} for more details).  Finally, our SFHs don't provide enough detail at young ages (less than $\simeq$ 1 Gyr) to seek features in the SFH that could be originated in the last LMC-SMC interaction, as predicted by \citet{Besla12}.

A determination of the LMC and SMC SFH, using data of similar depth as the one presented in this paper, but covering a really representative volume in both galaxies is necessary in order to obtain solid conclusions regarding the impact of galaxy interactions on the SFH of galaxies. It is possible that those improved and comprehensive SFHs, if coupled with a specific modelling of the impact of interactions on the SFH of the MCs, could be used to further constrain the orbits of the LMC-SMC-Milky Way system and as a consequence, to constrain the involved galactic masses.


\section{Summary and conclusions}

We have obtained the SFH of three fields in the outer LMC disk (R$_{gc}$= 4.0\degr\ -- 7.1\degr, or 3.5 to 6.2 Kpc), from wide field CMDs reaching the oMSTO with good photometric precision and high completeness. These SFHs  show common features, in the form of epochs of enhanced or decreased star forming activity, which intensity varies consistently and smoothly across the three fields, indicating a strong galactocentric stellar populations gradient in the sense that younger populations are more concentrated inwards. Two main epochs of star formation activity have been identified. The first, {\it old} star forming epoch, started at early times ($\simeq$ 13 Gyr ago) in all fields, and peaked around 10-8 Gyr ago. The second, {\it young} star forming epoch took place after a period of low activity occurred between 7 and 5--4 Gyr ago, and lasted till the present time, $\simeq$ 0.8 and 1.3 Gyr ago in fields LMC2, LMC1 and LMC0, respectively. The relative importance of the two epochs varies smoothly, from almost the same amount of stars formed in the two epochs in the case of the innermost field, to only 40\% of the stars formed in the more recent epoch for the outermost field. 

The variation of the age of the bulk of the youngest stellar population as a function of R$_{gc}$ is correlated with the currently measured HI column density. Under the reasonable assumption that stellar migration is not a dominant factor in shaping the distribution of the LMC stellar populations, this gradient implies that a displacement of the star formation activity towards smaller galactocentric radius occurred in the last $\simeq$ 1 Gyr. This could be related to a substantial shrinking of the HI disk able to form stars, either due to a reduction of pressure support as a result of gas depletion by star formation \citep{Stinson09} or to a removal of gas from the LMC disk by external agents such as ram-pressure from the Milky Way \citep{Mastropietro05}. It could also be the consequence of a compression of the gas disk as a result of ram-pressure from the Milky Way hot halo acting on the LMC's interstellar medium \citep{Mastropietro09}. In the latter cases, the process could be linked with the LMC entering the Milky Way's virial radius for the fist time \citep[e.g. as in one of the favoured models by][see their Figure 13]{Kallivayalil13}. These various theoretical possibilities imply substantially different azimuthal variations of the SFH, and thus a mapping of the SFH at different position angles and additional radius is key to get further insight onto the physical processes involved. This will be the subject of forthcoming papers in this series.

\section*{acknowledgments}

\small

We thank an anonymous referee for his/her careful review of the manuscript and many useful suggestions which have helped us to improve the paper. We are grateful to Dr. Staveley-Smith for providing  us with the HI column density measurements for our fields, and to Dr. V. Debattista, G. Besla, F. Pont, E. Hardy and R. Zinn, for useful discussions. The data presented in this paper were obtained as part of a joint project between the University of Chile and Yale University funded by the Fundaci\'on Andes. Finantial support for this work was provided by the IAC (grants P/310394 and P/301204) and the Education and Science Ministry of Spain (grant AYA2010-16717). 

\normalsize

\bibliography{LMC_bibtex}


\end{document}